\documentclass{emulateapj}

\usepackage{natbib}

\shorttitle{Accretion shocks in CTTSs: the role of local absorption on the X-ray emission}
\shortauthors{Bonito et al.}

\begin{document}

\title{Magnetohydrodynamic modeling of the accretion shocks in classical T Tauri stars: the
role of local absorption on the X-ray emission}

\author{R. Bonito\altaffilmark{1}}
\affil{Dip. di Fisica e Chimica, Universit\`a di Palermo, P.zza del Parlamento 1, 90134 Palermo, Italy}
\email{sbonito@astropa.unipa.it}

\author{S. Orlando}
\affil{INAF -- Osservatorio Astronomico di Palermo, P.zza del Parlamento 1,
90134 Palermo, Italy}

\author{C. Argiroffi\altaffilmark{1}}
\affil{Dip. di Fisica e Chimica, Universit\`a di Palermo, P.zza del Parlamento 1, 90134 Palermo, Italy}

\author{M. Miceli}
\affil{INAF -- Osservatorio Astronomico di Palermo, P.zza del Parlamento 1,
90134 Palermo, Italy}

\author{G. Peres\altaffilmark{1}}
\affil{Dip. di Fisica e Chimica, Universit\`a di Palermo, P.zza del Parlamento 1, 90134 Palermo, Italy}

\author{T. Matsakos\altaffilmark{2}}
\affil{Department of Astronomy and Astrophysics, The University of Chicago, Chicago IL 60637, USA}

\author{C. Stehle}
\affil{LERMA, Observatoire de Paris, Universit\'e Pierre et Marie Curie, Ecole Normale Superieure, Universite Cergy-Pontoise, CNRS, France}

\author{L. Ibgui}
\affil{LERMA, Observatoire de Paris, Universit\'e Pierre et Marie Curie, Ecole Normale Superieure, Universite Cergy-Pontoise, CNRS, France}

\altaffiltext{1}{INAF -- Osservatorio Astronomico di Palermo, P.zza del Parlamento 1, 90134 Palermo, Italy}

\altaffiltext{2}{LERMA, Observatoire de Paris, Universit\'e Pierre et Marie Curie, Ecole Normale Superieure, Universite Cergy-Pontoise, CNRS, France}

\begin{abstract}
We investigate the properties of X-ray emission from accretion shocks in classical T Tauri stars (CTTSs), generated where the infalling material impacts the stellar surface. Both observations and models of the accretion process reveal several aspects that are still unclear: the observed X-ray luminosity in accretion shocks is below the predicted value, and the density versus temperature structure of the shocked plasma, with increasing densities at higher temperature, deduced from the observations, is at odds with that proposed in the current picture of accretion shocks.
To address these open issues we investigate whether a correct treatment of the local absorption by the surrounding medium is crucial to explain the observations. 
To this end, we describe the impact of an accretion stream on a CTTS by considering a magnetohydrodynamic model. 
From the model results we synthesize the X-ray emission from the accretion shock by producing maps and spectra.
We perform density and temperature diagnostics on the synthetic spectra, and we directly compare the results with the observations.
Our model shows that the X-ray fluxes inferred from the emerging spectra are lower than expected because of the complex local absorption by the optically thick material of the chromosphere and of the unperturbed stream. Moreover, our model including the effects of local absorption explains in a natural way the apparently puzzling pattern of density versus temperature observed in the X-ray emission from accretion shocks.
\end{abstract}


\section{Introduction}
\label{Introduction}

Mass accretion processes onto classical T Tauri stars (CTTSs) generate shocks at the stellar surface as the accreting material impacts at supersonic infall velocities (typical values: $200 - 500$ km/s). The impacts generate slabs of post-shock plasma with temperature at a few millions degrees and high density ($n > 10^{11}$ cm$^{-3}$), emitting soft X-rays (\citealt{cg98}, \citealt{khs02}, \citealt{srn05}, \citealt{gls06}). Hydrodynamic and magneto-hydrodynamic (MHD) models of radiative accretion shocks support this scenario (e.g. \citealt{kur08, sao08, osa10, mcs13}, \citealt{oba13}). 
Recently, a further support arrived from spatially-resolved observations of impacts by dense fragments falling back in the Sun that show many analogies with accretion stream impacts (\citealt{rot13}).

The possibility to study the X-ray emission from accretion impacts in CTTSs is a fundamental tool to check the physical conditions of the material accreting from the disk onto the star. In fact, X-ray spectra of CTTSs, thanks to the density diagnostics offered by the He-like triplets, provide us with detailed information on the density and temperature structure of the hot post-shock region.

However, several aspects related to the X-ray emission from accretion shocks are still unclear and debated in the literature. 
In particular, the spectral analysis of the soft X-ray emission believed to originate from the impact region shows that, in all the cases for which high resolution spectra are available, the density derived from the O VII forbidden-to-intercombination ($f/i$) line ratio is always lower than that derived from the Ne IX $f/i$ ratio (\citealt{gls06}, \citealt{bcd10}). Considering that Ne IX originates from plasma hotter than that producing the O VII (4 MK vs ~2 MK, respectively), this result has been considered at first glance in contrast with 1D stationary models (\citealt{gsr07}) of post-shock region that indicate an increasing density and a decreasing temperature moving down in the post-shock region. Other scenarios have been proposed to interpret the observations, invoking the interaction between the accretion stream and the surrounding stellar corona \citep{bcd10}.

Another debated issue is the evidence that the mass accretion rates derived from X-rays are significantly lower than the mass accretion rates derived from other spectral bands (UV/optical/NIR observations; \citealt{cas11}), and that the observed X-ray luminosity in accretion shocks is, in general, well below the predicted value (\citealt{amp09}).

The local absorption, due to pre-shock material and surrounding chromosphere, is expected to play a fundamental role and could reconcile the above open issues.
A first attempt to investigate the theoretical observability of shock-heated accreting material in the X-ray band has been done by \citet{soa10} through 1D hydrodynamic modeling. These authors have pointed out the importance of the absorption from the optically thick material of the chromosphere in the post-shock plasma components that produce observable emission in the X-ray band. However, their analysis was based on 1D models and on a simple description of the absorption effect, preventing a detailed study of the observability of the X-ray emission from the impact region.

In this paper, we investigate the accretion shock properties in the X-ray band, properly taking into account geometry and absorption, with the aim to address the above open issues of the accretion theory. 
To this end, we consider a 2D MHD numerical model describing the impact of an accretion stream onto the surface of a CTTS. 
From the model results we synthesize the X-ray emission and explore the observability of the post-shock plasma. 
The X-ray emission is synthesized by considering the local absorption by optically thick plasma for different viewing angles of the impact region and for different wavelengths.
We focus on the emission line fluxes which traditionally are used for the density diagnostic in the observed spectra, namely the triplets of the He-like O~VII ($21.60, 21.80, 22.10 ~\AA$), Ne~IX ($13.45, 13.55, 13.70 ~\AA$), and Mg~XI ($9.17, 9.23, 9.31 ~\AA$) lines with the aim of providing useful diagnostics for a detailed and accurate interpretation of the observations.

\section{MHD modeling and synthesis of X-ray emission}
\label{model}

For our purposes, we performed 2D MHD simulations describing the impact of an accretion stream onto the chromosphere of a CTTS. We adopted the model described in \citet{osa10}. 
The stream impact is modeled by numerically solving the time-dependent MHD equations of mass, momentum, and energy conservation in a 2D cylindrical coordinate system $(r,z)$, assuming axisymmetry. The MHD model includes the gravity, the radiative cooling, and the magnetic-field-oriented thermal conduction (including the effects of heat flux saturation). The model also considers a detailed description of the stellar atmosphere, from the chromosphere to the transition region, and to the
corona. 
The MHD model is implemented using PLUTO (\citealt{mbm07}), a modular Godunov-type code for astrophysical plasmas. 

The star and accretion flow parameters of the simulations have been chosen in order to describe accretion stream impacts that are able to produce detectable X-ray emission (see also \citealt{soa10}).
The chosen set of parameters describes a typical X-ray emitting accretion stream as observed in MP Mus (\citealt{amp07}) or TW Hya (\citealt{khs02}).  
At odds with \citet{osa10}, we consider a stream characterized by a radial distribution of density (as obtained from 3D MHD simulations by \citealt{ruk04}), with $n_{\rm str} = 5\times 10^{11}$~cm$^{-3}$ at the center of the stream and $n_{\rm str} = 5\times 10^{10}$~cm$^{-3}$ at the border, propagating through a uniform stellar magnetic field with strength $B = 500$ G oriented along the $z$ axis. Initially the stream is in pressure equilibrium with the stellar corona and has a circular cross-section with a radius $r_\mathrm{str} = 10^{10}$~cm. We follow the stream evolution for about $3000$ s.

From the evolution of the temperature, density, and velocity of the plasma in the 2D spatial domain, we reconstruct the 3D spatial distributions by rotating the 2D slab around the symmetry $z$ axis. From the values of emission measure, $EM$, and temperature in each computational cell, we synthesize the corresponding emission using the CHIANTI atomic database (\citealt{ldy12}) and assuming metal abundances of $0.5$ of the solar values (as deduced from X-ray observations of CTTSs; \citealt{tgb07}). The spectral synthesis takes into account the Doppler shift of lines due to the component of plasma velocity along the line of sight (LoS).
The local absorption is accounted for by computing the X-ray spectrum from each cell and by filtering it through the absorption column density along the LoS.
The absorption is computed using the absorption cross-sections as a function of wavelength from \cite{bm92}. Note that the bulk of the absorption originates within relatively cold material, given that the soft X-ray opacity drops at high temperature ($T>10^6$ K), as shown by \cite{kk84}. The local absorption originates from the optically thick material distributed around the post-shock region belonging to the unperturbed stream above the hot slab and to the perturbed chromosphere that may surround the slab. 
By integrating the absorbed X-ray spectra from the cells in the whole spatial domain, we derive X-ray images and spectra emerging from the impact region. 
We computed the emerging spectrum at different times, in order to take into account the intrinsic variability of the post-shock region.
In this context, we subtract the emission from the coronal component and do not consider the absorption due to the interstellar medium. 

In the following, the synthetic maps and spectra of X-ray emission
are used to explore the effects of the local absorption on the observability of the X-ray emitting plasma and the effects of the geometry of the system by exploring different points of view at which the impact region is observed.

\section{Results}
\label{results}

The evolution of the accretion impact and of the post-shock plasma are shown in the movie provided as online material showing the 2D spatial distributions of mass density and temperature.
The simulation shows that the accreting material flows along the magnetic field lines and a hot slab is generated at the base of the accretion column.
During the whole evolution, the plasma beta is $< 1$ in the stream and, in particular, in the shocked slab. Thus the magnetic field prevents mass and energy exchange across field lines and the stream is structured as a bundle of fibrils, each independent on the others.
A detaild study of the physical properties of the fibrils is discussed in \cite{mcs13}. In each fibril, the shock position oscillates due to intense radiative cooling at the base of the slab, with alternating phases of expansion and collapse of the post-shock region. During the expansion phase the post-shock temperature is higher than during the collapse (e.g. see Fig. 3 of \citealt{oba13}). The slightly different mass density of the fibrils results in different instability periods, as also by random phases of the oscillations after few cycles.

\begin{figure}[!t]
\centerline{\includegraphics[width=\columnwidth,clip]{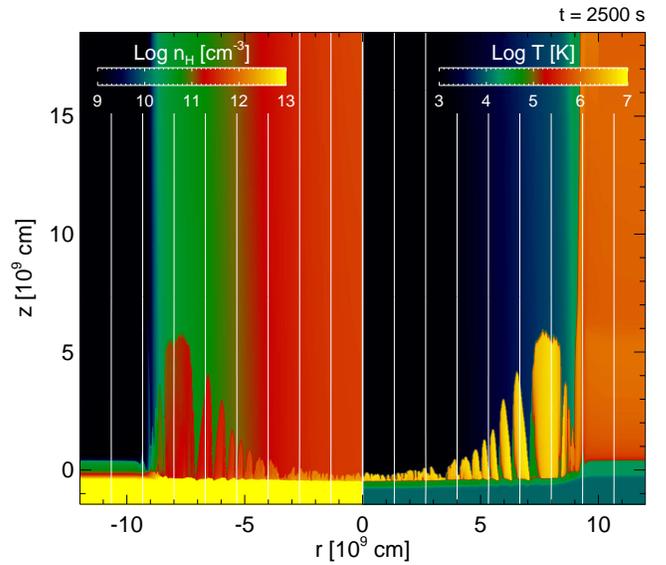}}
\caption{Density (on the left) and temperature (on the right) 2D map as derived from the model at $2500$ s.}
\label{2D}
\end{figure}

As the stream density varies radially, the stand-off height of the hot slab\footnote{Which depends on the pre-shock density (\citealt{soa10}).} is minimum (maximum) at the stream center (border). Fig. \ref{2D} shows maps of density (left panel) and temperature (right panel) at the labeled time.

\begin{figure}[!t]
\centerline{\includegraphics[width=\columnwidth,clip]{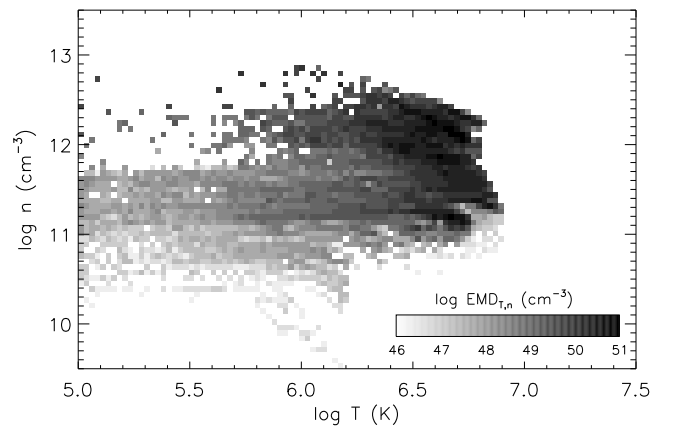}}
\caption{Emission measure distribution as a function of temperature and density derived from the model at $2500$ s of evolution.}
\label{n-T}    
\end{figure}

We derived the emission measure distribution as a function of temperature and density from the 2D model (Fig. \ref{n-T}). 
It is evident that the simplistic scenario of a plane-parallel stratified structure fails to be valid as a complex distribution of density and temperature of the post-shock region is found. Our detailed numerical model reveals the complexity of the shock and post-shock region.

\begin{figure*}[!t]
\centerline{\includegraphics[width=9cm,clip]{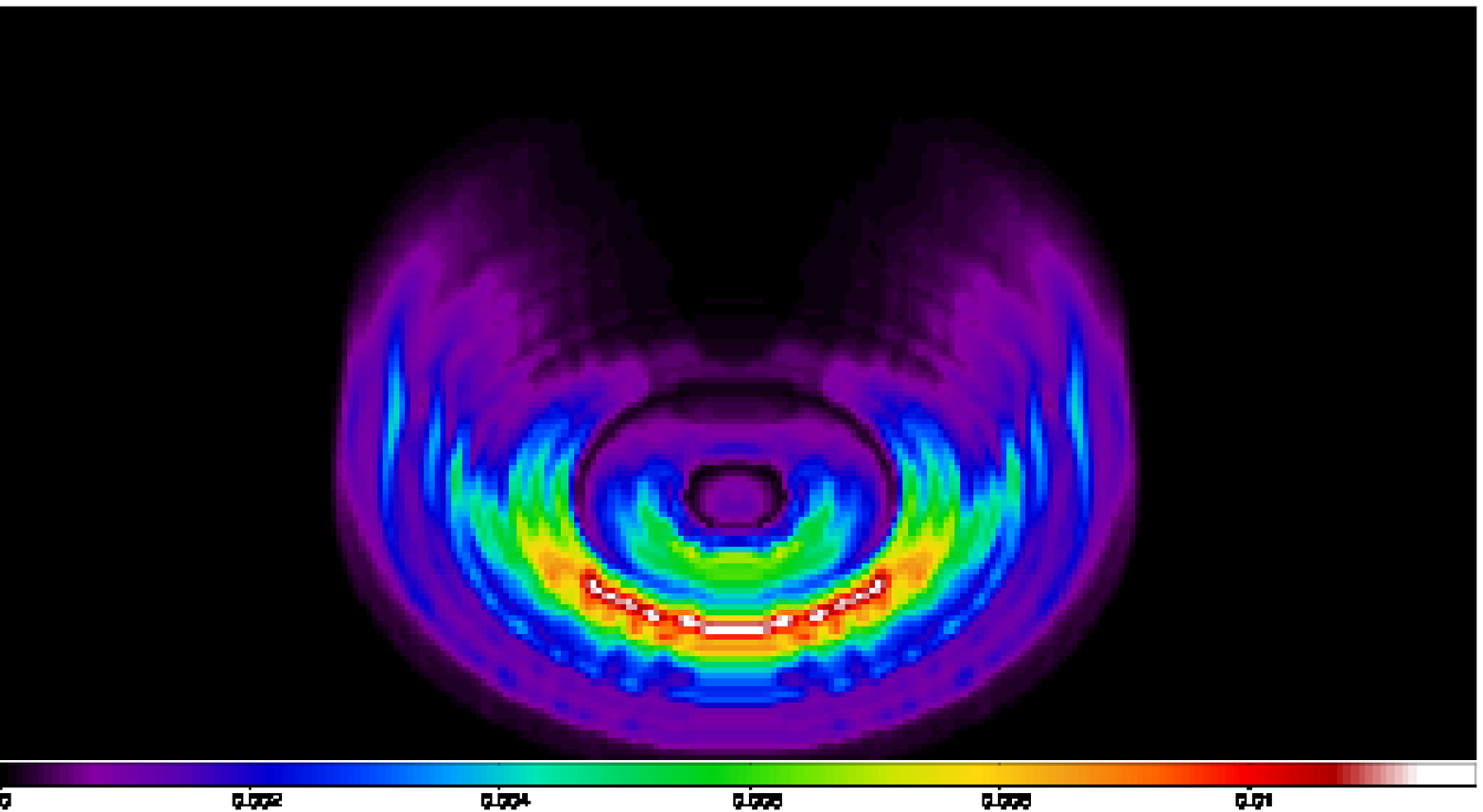}\hspace{0.2cm}\includegraphics[width=9cm,clip]{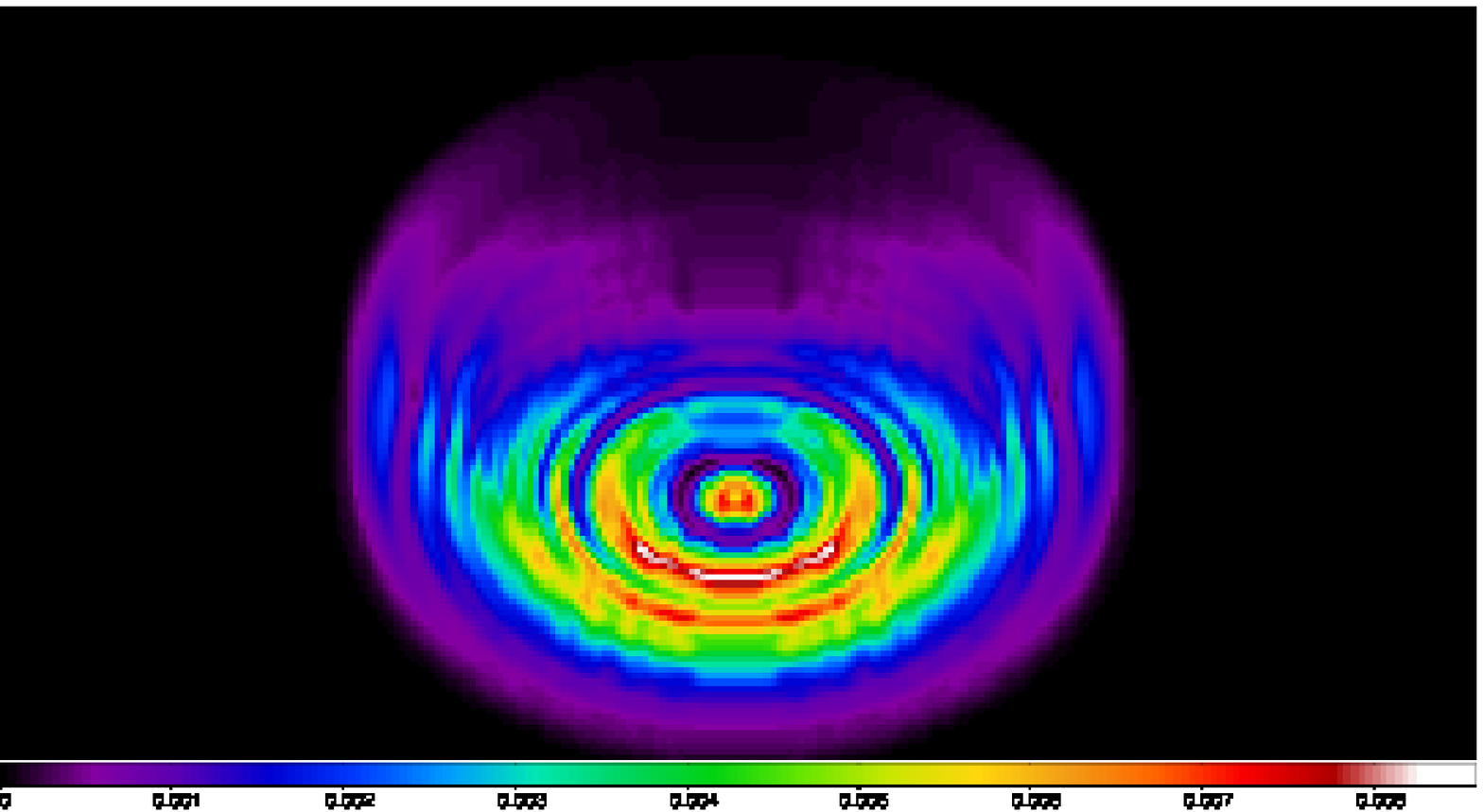}}
\centerline{\includegraphics[width=9cm,clip]{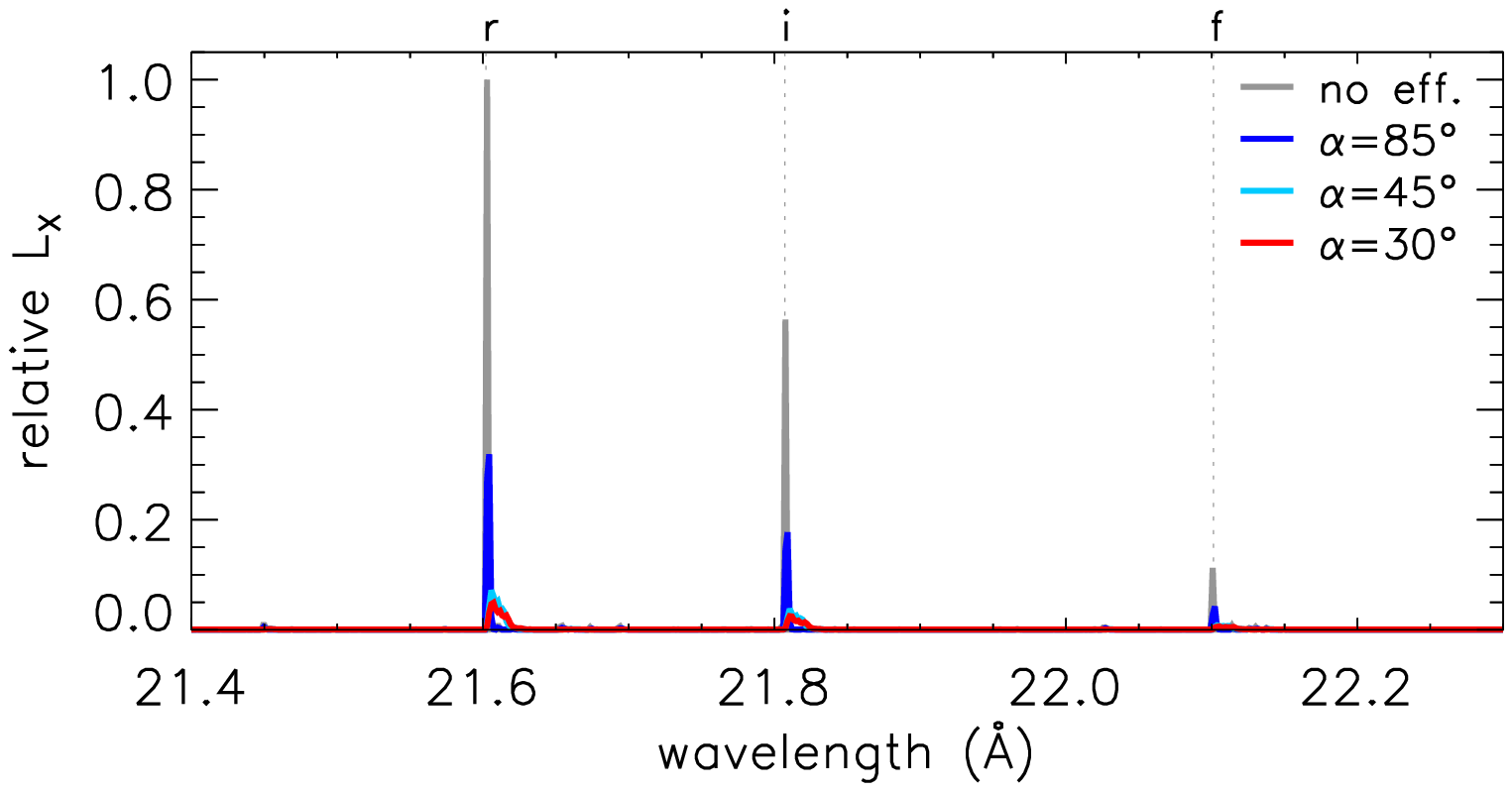}\hspace{0.1cm}\includegraphics[width=9cm,clip]{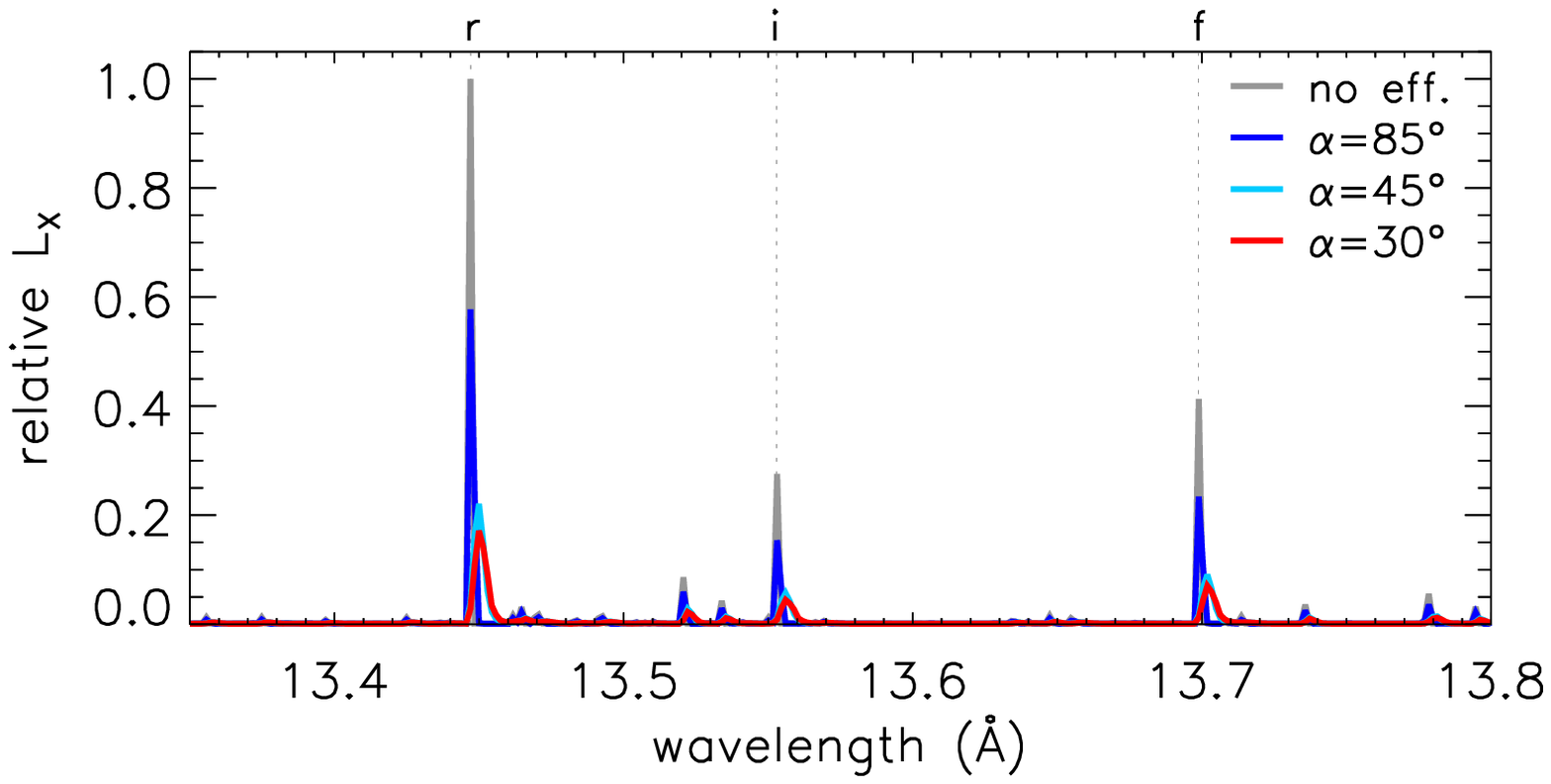}}
\caption{Upper panels: enlargement of the X-ray emitting impact region. The emission is synthesized from the model at $2500$ s for O VII (left panel) and Ne IX (right panel); the LoS of the observer forms an angle of $\alpha = 45^{\circ}$ with respect to the stream axis.
Lower panels: spectral region near the O VII (left panel) and Ne~IX (right panel) triplets derived from the model. Each spectrum (in units of ph/s/bin; 
blue line: $\alpha = 85^{\circ}$, cyan line: $\alpha = 45^{\circ}$, red line: $\alpha = 30^{\circ}$) is divided by the maximum value of the unabsorbed spectrum (gray line) in the plotted range. The effects of local absorption and Doppler shift are taken into account.}
\label{emiss}
\end{figure*}

We derived synthetic maps and spectra in the X-ray band from the model. 
All the fibrils contribute to the X-ray emission. The resulting spectrum integrates the contribution of the independent fibrils with different densities and temperatures that compete to the phase of oscillation.
Fig. \ref{emiss} shows maps of X-ray emission derived from the model relative to the O VII and the Ne IX triplets emission (upper left and right panels, respectively) at an angle between the LoS and the accretion stream $\alpha = 45^{\circ}$ and with the effects of local absorption and Doppler shift taken into account.
The synthetic maps show that the local absorption from the accretion stream heavily obscures the central (denser) part of the accretion shock in the O VII band while in the Ne IX band, where the effects of the absorption are smaller, we can see deeper inside the accretion stream.  
Fig. \ref{emiss} (lower panels) shows the spectral region near the O VII (left panel) and Ne~IX (right panel) triplets averaging over different frames (between $2000 - 2500$ s). 
When the absorption is taken into account, the flux of the lines can be heavily reduced. The amount of reduction depends on the angle $\alpha$ and is due mainly to the dense pre-shock accretion stream. The effects of absorption appear to be dominant for $\alpha = 30^{\circ}$ (see Fig. \ref{emiss}, lower panels).
We investigated how accurately the unabsorbed X-ray luminosity of the accretion shock can be inferred from the emerging X-ray spectrum. To this aim we derived the absorbing column $N_{\rm H}$ by the observed ratio of the O~VII resonance line at $21.60 ~\AA$ to the O~VII line at $18.63 ~\AA$, and starting from this $N_{\rm H}$ we computed the unabsorbed flux of the O~VII resonance line. We obtained that the unabsorbed flux was underestimated with respect to the correct value by a factor of $2$ in the case of $\alpha=30^{\circ}$ ($1.8$ and $1.5$ for $\alpha=45^{\circ}$ and $85^{\circ}$, respectively). Hence the {\it average} $N_{\rm H}$ value inferred from the emerging spectrum is underestimated, likely because the most absorbed portions of the post shock region do not contribute significantly to the emerging spectrum.

\begin{figure}[!t]
\centerline{\includegraphics[width=\columnwidth,clip]{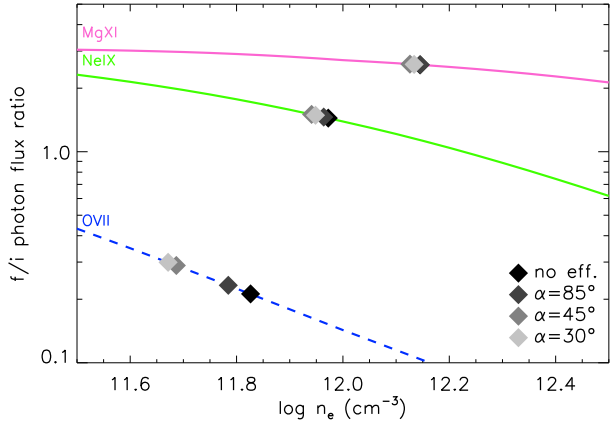}}
\caption{Ratios of forbidden-to-intercombination line fluxes as functions of density for several He-like lines (O, Ne, and Mg) with the values (diamond) derived from our analysis.}
\label{fi}      
\end{figure}

We used the total synthetic spectra to investigate the evidence that the shocked plasma appears denser at higher temperatures (\citealt{bcd10}). To this end, we apply to the spectra the same density diagnostic techniques commonly used to analyze observed spectra and based on the $f/i$ flux ratio derived for several He-like lines (\citealt{gj69}; \citealt{pmd01}). Fig. \ref{fi} shows the $f/i$ line fluxes as a function of density for several He-like triplets (O, Ne, and Mg) synthesized from the model. 

We find that plasma components of the shocked slab with higher temperatures are also characterized by higher densities, in full agreement with the observations, and in particular this trend is amplified when the local absorption effect is taken into account. 
In fact, when a 2D MHD model of the stream impact is applied, as the one developed by us in this case, the complexity of the emitting region is evident. 
In particular, the distribution of density vs. temperature is far from that expected for a plane-parallel stratified structure of the post-shock region. Furthermore, the local absorption prevents us from the observation of deeper and denser regions for more absorbed emitting regions. This effect is different for the different triplets inspected.
In fact, the softer part of the spectrum suffers a higher absorption: the O~VII triplet, which originates at lower temperature than the Ne~IX triplet, is more absorbed and then in the wavelengths of the O~VII triplet we cannot observe as deep as we can in Ne~IX energy band (compare upper left and right panels of Fig. \ref{emiss}). Therefore, when the absorption is taken into account, the Ne~IX emitting plasma appears denser (as we can observe deeper) than the O~VII emitting region (see Fig. \ref{fi}).

\section{Discussions and conclusions}
\label{discussions}

The X-ray emission from accretion shocks detected in CTTSs opens several issues in both observations and models.
In particular, the evidence that the X-ray luminosity in accretion shocks is below the predicted value and that the density vs. temperature structure of the shocked plasma is not in agreement with the simplified picture of the post-shock region urges us to investigate whether a correct treatment of the local absorption by the surrounding medium can significantly affect the observed X-ray emission.

To address these issues, we consider a 2D MHD model describing accretion stream impacts in CTTSs. We explore the effect of a density-structured stream on the emerging X-ray emission.
We synthesize the X-ray emission of the post-shock plasma from the model results including the effects of the local absorption and by taking into account that the contribution to the absorption is different for different components of the hot post-shock medium, for different inclination angles, and for different inspected wavelengths.

We find that the X-ray emission originating in the shocked slab can be heavily reduced due to the absorption by the optically thick material surrounding the slab. 
Moreover different regions of the post-shock suffer very different extinctions, as shown in Fig. \ref{emiss} (upper panels), with some regions being completely obscured. As a consequence any average extinction derived from the data, and hence the amount of emitting plasma, could be significantly underestimated.

This has a direct implication on the derivation of the mass accretion rate in the X-ray band that is expected to be largely underestimated, in agreement with observations. At the same time, the point of view from which the impact region is observed plays a crucial role because of the different distribution of the optically thick material along the LoS.

Our results demonstrate that it is necessary to adopt an appropriate description of the local absorption together with a realistic model which properly describes the accretion shock to reveal the complexity of the emitting region.
We verified that the post-shock plasma presents a complex distribution of density and temperature. We also found that the observation of deeper and denser regions is hampered in the soft X-rays (e.g. at the energies of the low temperature O~VII triplet) because of the local absorption, thus explaining why the cooler plasma appears also more tenuous.
By directly comparing our model results with observations, we confirm that the observed evidence that components with higher temperatures are also characterized by higher densities, a trend so far not explained, can be explained in a natural way by a detailed and realistic model.

\begin{acknowledgements}
Pluto is developed at the Turin Astronomical Observatory in collaboration
with the Department of General Physics of Turin University. We acknowledge
the CINECA Award N. HP10BG6HA5,2012 for the availability of high
performance computing resources and support, and the computer resources,
technical expertise and assistance provided by the Red Espanola de
Supercomputacion (award N. AECT-2012-2-0001).
T.M., L.I. and C.S. acknowledge the support of the ANR STARSHOCK project ANR-08-BLAN-0263-07-2009/2013. T.M. was in part supported by NASA ATP grant NNX13AH56G.  L.I. and C.S. acknowledge the support of Labex Plas@par (ANR-11-IDEX-0004-02)
\end{acknowledgements}

\bibliographystyle{apj}
%\bibliography{references}

\end{document}